\documentclass{ws-ijgmmp}

\begin{document}

\markboth{A. Paliathanasis, M. Tsamparlis}
{The geometric origin of Lie point symmetries of the Schr\"{o}dinger
and the Klein Gordon equations}

%\catchline{}{}{}{}{}
%
%%%%%%%%%%%%%%%%%%%%%%%%%%%%%%%%%%%%%%%%%%%%%%%%%%%%%%%%%%%%%%%%%%%%

\title{The geometric origin of Lie point symmetries of the Schr\"{o}dinger
and the Klein Gordon equations}

\author{ANDRONIKOS PALIATHANASIS}

\address{Faculty of Physics, Department of Astrophysics - Astronomy -
Mechanics,University of Athens, Panepistemiopolis, Athens 157 83,
Greece\\
\email{anpaliat@phys.uoa.gr}}

\author{MICHAEL TSAMPARLIS}

\address{Faculty of Physics, Department of Astrophysics - Astronomy -
Mechanics,University of Athens, Panepistemiopolis, Athens 157 83,
Greece\\
\email{mtsampa@phys.uoa.gr}}

\maketitle

\begin{history}
%\received{(Day Month Year)}
%\revised{(Day Month Year)}
\end{history}

\begin{abstract}
We determine the Lie point symmetries of the Schr\"{o}dinger and
the Klein Gordon equations in a general Riemannian space. It is shown that these
symmetries are related with the homothetic and the conformal algebra of the
metric of the space respectively. We consider the kinematic metric defined by
the classical Lagrangian and show how the Lie point symmetries of the
Schr\"{o}dinger equation and the Klein Gordon equation are related with the
Noether point symmetries of this Lagrangian. The general results are applied
to two practical problems a. The classification of all two and three
dimensional potentials in a Euclidian space for which the Schr\"{o}dinger
equation and the Klein Gordon equation admit Lie point symmetries and b. The
application of Lie point symmetries of the Klein Gordon equation in the
exterior Schwarzschild spacetime and the determination of the metric by means
of conformally related Lagrangians.

\end{abstract}


\section{Introduction}

A systematic method to facilitate the solution of differential equations (DE)
is the use of Lie symmetries, because the latter provide the first order
invariants which can be used to reduce the DE. The Schr\"{o}dinger equation
and the Klein Gordon equation are two important equations of Quantum Physics.
Therefore it is important that we determine their Lie point symmetries and use
them either in order to find invariant solutions\ using Lie symmetry methods
\cite{StephaniB,Ibragimov} or to obtain first integrals which will ease the
search for analytic solutions.

Nowadays the determination of Lie symmetries of DEs can be done mechanically
by symbolic computer programs, for instance CRACK, MATHLIE, DIMSYM, SADE and
others (see \cite{Wolf,Baumann,PrinceCP,SADE} and references therein). However
this option is prohibited when one works in Riemannian spaces and in higher
dimensions. Therefore it is important that one finds geometric arguments which
provide the Lie symmetries of DEs in general Riemannian spaces irrespectively
of the (finite) dimension f the space. Indeed one finds in the recent
literature (\cite{Aminova95,Aminova00,Prince,Qadir,2DPot,Bozhkov,JGP}) which
provide the Lie symmetry vectors directly from the collineations of the metric.

The Lie symmetries of the heat equation and the Poisson equation in a general
Riemannian space have been determined in previous works \cite{Bozhkov,JGP}. We
use these results to find the Lie symmetries of Schr\"{o}dinger and the Klein
Gordon equation in a general Riemannian space.

An important element of the present study is the concept of conformally
related Lagrangians, that is Lagrangians which under a conformal
transformation of the metric and the potential lead to the same equations but
for different dynamic variables. The condition for this is that the
Hamiltonian vanishes. Because the dynamic variables of these Lagrangians are
not the standard ones in general the Hamiltonian is not relevant to the energy
of the system.

From each Lagrangian describing a dynamical system we define a metric, called
the kinematic metric, characteristic to the dynamical system described by this
Lagrangian. As it will be shown the conformal symmetries of this metric are in
close relation with the Noether point symmetries of the equations of motion.
Furthermore the kinetic metric of the Lagrangian defines the Laplace operator,
hence the corresponding Klein Gordon equation whose Lie symmetries are
consequently expressed in terms of the conformal symmetries of the kinematic metric.

The structure of the paper is as follows.{\Large \ }In section \ref{Col1} we
give a brief account of the basic properties of Lie point symmetries of
differential equations and the conformal algebra of a Riemannian space. In
section \ref{Symmetries of Lagrangian} we consider the classical Lagrangian
\[
L\left(  x^{k},\dot{x}^{k}\right)  =T\left(  x^{k},\dot{x}^{k}\right)
-V\left(  x^{k}\right)  ~,~\dot{x}^{k}=\frac{dx^{k}}{dt}%
\]
of a particle moving in a Riemannian space under the action of the
potential~$V\left(  x^{k}\right)  $, where $T\left(  x^{k},\dot{x}^{k}\right)
$ is the Kinetic energy defined as follows $T=\frac{1}{2}g_{ij}\left(
x^{k}\right)  \dot{x}^{i}\dot{x}^{j}$ and $g_{ij}$ is the metric of the
Riemannian space. We recall some basic results concerning the relation of
Noether point symmetries of the Lagrangian with the homothetic algebra of the
metric $g_{ij}.$ Then we show that the Noether symmetries of two conformally
related Lagrangians are generated from the common conformal algebra.

In section \ref{LiePointSchKG} we study the Lie point symmetries of the
Schr\"{o}dinger and the Klein Gordon equations in a general Riemannian space
and show that the Lie symmetries are generated from the conformal algebra of
the metric which defines the Laplace operator. Using the geometric character
of Noether symmetries for the classical Lagrangian and that of\ the Lie point
symmetries of the Schr\"{o}dinger and of the Klein Gordon equation we
establish the connection between the two. More specifically it is shown that
if an element of the homothetic algebra of the kinetic metric generates a
Noether point symmetry for the classical Lagrangian then it also generates a
Lie point symmetry for the Schr\"{o}dinger equation.

In section \ref{Symmetries of the Lagrangian with non constant gauge function}
we consider the case of Noether point symmetries of the classical Lagrangian
which are generated from a gradient Killing vector (KV) or a gradient
homothetic vector (HV) and show that the Lie symmetry in both cases is a
non-local symmetry of the Klein Gordon equation.

In sections \ref{Classification} and \ref{StaticSS}, we demonstrate the use of
the previous general results to various interesting practical situations. In
particular in section \ref{Classification} we find all two and three
dimensional potentials in a Euclidian space for which the Schr\"{o}dinger and
the Klein Gordon equations admit non trivial Lie point symmetries. In section
\ref{StaticSS} we apply the results of section \ref{LiePointSchKG} in order to
derive invariant solutions of the Wheeler--DeWitt equation\ and solve the
field equations by means of conformally related Lagrangians in a static
spherical symmetric spacetime. Finally in section \ref{conclusion} we close
with the conclusion.

\section{Symmetries of DEs and collineations of Riemannian spaces}

\label{Col1}

For the convenience of the reader and the completeness of the paper we state
briefly some basic results concerning the Lie point symmetries of differential
equations and the conformal algebra of a Riemannian space.

\subsection{Lie symmetries of DEs}

A partial differential equation (PDE) is a function $H=H\left(  x^{i}%
,u,u_{,i},u_{,ij},..\right)  $ in the jet space $\bar{B}_{\bar{M}},~$where
$x^{i}$ are the independent variables and $u^{A}$ are the dependent variables.
The infinitesimal transformation%
\begin{align}
\bar{x}^{i}  & =x^{i}+\varepsilon\xi^{i}\left(  x^{k},u\right) \label{pde1}\\
\bar{u}  & =\bar{u}+\varepsilon\eta\left(  x^{k},u\right) \label{pde2}%
\end{align}
where $\varepsilon$ is an infinitesimal parameter, is a Lie point symmetry of
$H$ with generator%
\begin{equation}
X=\xi^{i}\left(  x^{k},u^{B}\right)  \partial_{t}+\eta^{A}\left(  x^{k}%
,u^{B}\right)  \partial_{B}\label{pde3}%
\end{equation}
if there exist a function $\lambda$ such that the following condition holds
\cite{StephaniB,Ibragimov}%
\begin{equation}
X^{\left[  n\right]  }H\left(  x^{i},u,u_{,i},u_{,ij},..\right)  =\lambda
H\left(  x^{i},u,u_{,i},u_{,ij},..\right)  ~,modH=0.\label{pde4}%
\end{equation}
$X^{\left[  n\right]  }$ is the nth prolongation of (\ref{pde3}) defined as
follows%
\[
X^{\left[  n\right]  }=X+\eta_{\left[  i\right]  }\partial_{u_{i}}%
+...+\eta_{\left[  ij...i_{n}\right]  }\partial_{u_{ij...i_{n}}}%
\]
where%
\[
\eta_{\left[  i\right]  }=D_{i}\eta-u_{,j}D_{i}\xi^{j}%
\]%
\[
\eta_{\left[  ij..i_{n}\right]  }=D_{i_{n}}\eta_{\left[  ij..i_{n-1}\right]
}-u_{ij..k}D_{i_{n}}\xi^{k}.
\]

The Lie point symmetry vector field $\mathbf{X}=\xi\left(  t,x^{k}\right)
\partial_{t}+\eta^{i}\left(  t,x^{k}\right)  \partial_{x^{i}}$ of the field
equations resulting from the Lagrangian function $L=L\left(  t,x^{k},\dot
{x}^{k}\right)  ,$ is a Noether point symmetry of these equations if there
exist a function $f=f\left(  t,x^{k}\right)  $ so that the following condition
holds \cite{StephaniB}
\begin{equation}
\mathbf{X}^{\left[  1\right]  }L+\frac{d\xi}{dt}L=\frac{df}{dt}%
.\label{NSCS.1a}%
\end{equation}
$\mathbf{X}^{\left[  1\right]  }$ is the first prolongation of $\mathbf{X}$.
The corresponding Noether integral is
\begin{equation}
I=\xi\left(  \frac{\partial L}{\partial\dot{x}^{k}}\dot{x}^{k}-L\right)
-\eta^{i}\frac{\partial L}{\partial x^{i}}+f.
\end{equation}

\subsection{Collineations of Riemannian spaces}

In the following $L_{\xi}$ denotes Lie derivative with respect to the vector
field $\xi^{i}$. A vector field $\xi^{i}$ is a CKV of a metric $g_{ij}$ if
\[
L_{\xi}g_{ij}=2\psi g_{ij}%
\]
where $L_{\xi}$ denotes Lie derivative with respect to the vector field
$\xi^{i}$. If $\psi=0$ then $\xi^{i}$ is a Killing vector (KV), if $\psi
_{,i}=0$ then $\xi^{i}$ is a homothetic vector (HV)~and if $\psi_{;ij}%
=0$,~$\xi^{i}$ is a special conformal Killing vector (sp.CKV). If a CKV is a
not a KV/HV or sp.CKV is called a proper CKV~\cite{Katzin}.

Two metrics $g_{ij},~\bar{g}_{ij}$ are conformally related if $\bar{g}%
_{ij}=N^{2}g_{ij}$ where the function ~$N^{2}$ is the conformal factor. If
$\xi^{i}$ is a CKV\ of the metric $\bar{g}_{ij}$ so that $L_{\xi}\bar{g}%
_{ij}=2\bar{\psi}\bar{g}_{ij}$ then $\xi^{i}$ it is also a CKV of the metric
$g_{ij}$, that is, $L_{\xi}g_{ij}=2\psi g_{ij}$ where the conformal factor
\[
\psi=\bar{\psi}N^{2}-NN_{,i}\xi^{i}.
\]
The last relation implies that two conformally related metrics have the same
conformal algebra but with Killing/homothetic/special conformal subalgebras
spanned by different vector fields. For example a KV for one may be proper CKV
for the other. This is an important observation which shall be useful in the
following sections.

\section{Conformal Lagrangians and Noether symmetries}

\label{Symmetries of Lagrangian}

In this section we examine the construction of Noether point symmetries of the
classical Lagrangian and the transition of these symmetries to conformally
related Lagrangians.

Consider the Lagrangian of a particle moving under the action of a potential
$V(x^{k})$ in a Riemannian space with metric $g_{ij}$
\begin{equation}
L=\frac{1}{2}g_{ij}\dot{x}^{i}\dot{x}^{j}-V\left(  x^{k}\right) \label{CLN.05}%
\end{equation}
where $\dot{x}=\frac{dx}{dt}.~$The equations of motion follow from the action%
\begin{equation}
S=\int dt\left(  L\left(  x^{k},\dot{x}^{k}\right)  \right)  =\int dt\left(
\frac{1}{2}g_{ij}\dot{x}^{i}\dot{x}^{j}-V\left(  x^{k}\right)  \right)  .
\end{equation}
Consider the change of variable $t\rightarrow\tau$ defined by the requirement%
\begin{equation}
d\tau=N^{2}\left(  x^{i}\right)  dt.
\end{equation}
In the new coordinates $(\tau,x^{i}),$ the action becomes%
\begin{equation}
S=\int\frac{d\tau}{N^{2}\left(  x^{k}\right)  }\left(  \frac{1}{2}g_{ij}%
N^{4}\left(  x^{k}\right)  x^{\prime i}x^{\prime j}-V\left(  x^{k}\right)
\right)  ~~
\end{equation}
where $x^{\prime i}=\frac{dx}{d\tau}$ and the Lagrangian is transformed to the
new Lagrangian%
\begin{equation}
\bar{L}\left(  x^{k},x^{\prime k}\right)  =\frac{1}{2}N^{2}\left(
x^{k}\right)  g_{ij}x^{\prime i}x^{\prime j}-\frac{V\left(  x^{k}\right)
}{N^{2}\left(  x^{k}\right)  }.
\end{equation}
If we consider a conformal transformation (not a coordinate transformation!)
of the metric $\bar{g}_{ij}=N^{2}\left(  x^{k}\right)  g_{ij}$ and a new
potential function $\bar{V}\left(  x^{k}\right)  =\frac{V\left(  x^{k}\right)
}{N^{2}\left(  x^{k}\right)  },$ then, the new Lagrangian $\bar{L}\left(
x^{k},x^{\prime k}\right)  $ in the new coordinates $\tau,x^{k}$ takes the
following form,%
\begin{equation}
\bar{L}\left(  x^{k},x^{\prime k}\right)  =\frac{1}{2}\bar{g}_{ij}x^{\prime
i}x^{\prime j}-\bar{V}\left(  x^{k}\right) \label{CLN.09}%
\end{equation}
which is of the same form as the Lagrangian $L$ in equation (\ref{CLN.05}).
Lagrangians $L\left(  x^{k},\dot{x}^{k}\right)  $ and $\bar{L}\left(
x^{k},x^{\prime k}\right)  $ will be called conformal. In this framework, the
action remains the same, i.e. it is invariant under the change of parameter
and the equations of motion in the new variables $(\tau,x^{i})$ will be the
same with the equations of motion for the Lagrangian $L$ in the original
coordinates $(t,x^{k})$ .

In \cite{2DPot}, it was shown that the Noether symmetries of a Lagrangian of
the form (\ref{CLN.05}) follow from the homothetic algebra of the metric
$g_{ij}$. The same applies to the Lagrangian $\bar{L}\left(  x^{k},x^{\prime
k}\right)  $ and the metric $\bar{g}_{ij}.$~The conformal algebra of the
metrics $g_{ij},\bar{g}_{ij}$ are (as a set) the same but their closed
subgroups of HVs and KVs are spanned (in general) by different vector
fields~\cite{Yano}. This observation leads us to the conclusion, that the
Noether point symmetries of the conformally related Lagrangians $L\left(
x^{k},\dot{x}^{k}\right)  $ and $\bar{L}\left(  x^{k},x^{\prime k}\right)  $
are contained in the common conformal algebra of the metrics $g_{ij},\bar
{g}_{ij}~$\cite{TsamCap}.

In the following section using the general results of \cite{JGP} we study the
Lie point symmetries of Schr\"{o}dinger and of the Klein
Gordon equation in Riemannian manifolds.

\section{Lie symmetries of Schr\"{o}dinger and the Klein Gordon equation}

\label{LiePointSchKG}

Consider the Schr\"{o}dinger equation \footnote{We absorb the constant
$\hslash$ and the imaginary unit $i$, in the variables $x^{k},t$ respectively}
(linear diffusion equation)
\begin{equation}
\Delta u-u_{,t}=V\left(  x^{i}\right)  u\label{lsk.01}%
\end{equation}
and the Klein Gordon equation%

\begin{equation}
\Delta u=V\left(  x^{i}\right)  u\label{lsk.02}%
\end{equation}
in a Riemannian space, where $\Delta$ is the Laplace operator of the
Riemannian space with metric $g_{ij}\left(  x^{k}\right)  ~$defined as
follows
\begin{equation}
\Delta u=\frac{1}{\sqrt{\left\vert g\right\vert }}\frac{\partial}{\partial
x^{i}}\left(  \sqrt{\left\vert g\right\vert }g^{ij}\frac{\partial}{\partial
x^{j}}\right)  u=g^{ij}u_{,ij}-\Gamma^{i}u_{,i}.\label{lsk.03}%
\end{equation}
$\Gamma_{jk}^{i}\left( x^{k}\right)  $ are the Christofell symbols of the
metric $g_{ij}$ and $\Gamma^{i}=g^{jk}\Gamma_{jk}^{i}$. In \cite{JGP} it has
been shown that the Lie point symmetries of the second order PDE
\begin{equation}
A^{ab}\left(  x^{c}\right)  u_{,ab}-F\left(  x^{c},u,u_{,c}\right)
=0\label{lsk.04}%
\end{equation}
are generated from the elements of the conformal algebra of the second rank
tensor~$A^{ab}\left(  x^{c}\right)  $. \ Equations (\ref{lsk.01}) and
(\ref{lsk.02}) are special cases of (\ref{lsk.04}), that is, the Lie point
symmetries of the Schr\"{o}dinger equation and those of the Klein Gordon
equation are generated from the conformal algebra of the metric $g_{ij}$.
Therefore by applying the general results of \cite{JGP} we have the following
propositions which relate the Lie point symmetries of equations (\ref{lsk.01})
and (\ref{lsk.02}) with the geometry of the underlying Riemannian space which
defines the Laplace operator.

\begin{proposition}
[Lie Symmetries of Schr\"{o}dinger equation]The Lie point symmetries of
Schr\"{o}dinger equation (\ref{lsk.01}) are generated from the elements of the
homothetic algebra of the metric $g_{ij}$ as follows.\newline a) $Y^{i}$ is a
non-gradient HV/KV. \ The generic Lie symmetry vector is
\begin{equation}
X=\left(  2c\psi t+c_{1}\right)  \partial_{t}+cY^{i}\partial_{i}+\left(
a_{0}u+b\left(  t,x\right)  \right)  \partial_{u}%
\end{equation}
with constraint equation%
\begin{equation}
H\left(  b\right)  -bV=0~,~~cL_{Y}V+2\psi cV+a_{0}=0.
\end{equation}
where $b\left(  t,x^{k}\right)  $ satisfies (\ref{lsk.01}).\newline b)
$Y^{i}=H^{,i}$ is a gradient HV/KV. The generic Lie symmetry vector is%
\[
X=\left(  2\psi\int Tdt\right)  \partial_{t}+TS^{,i}\partial_{i}+\left(
\left(  -\frac{1}{2}T_{,t}S+F\left(  t\right)  \right)  u\right)  \partial_{u}%
\]
with constraint equation%
\begin{equation}
L_{H}V+2\psi V-\frac{1}{2}c^{2}H+d=0
\end{equation}
and the functions $T,F$ satisfies the following system
\begin{equation}
T_{,tt}=c^{2}T~,~\frac{1}{2}T_{,t}\psi+F_{,t}=Td.
\end{equation}

\end{proposition}

Similarly \ for the Klein Gordon equation~(\ref{lsk.02}) we have the following result

\begin{proposition}
[Lie symmetries of Klein Gordon equation]\label{KG1}The Lie point symmetries
of Klein Gordon equation (\ref{lsk.02})$\ $are generated from the CKVs of the
metric~$g_{ij},$ defining the Laplace operator, as follows\newline a) for
$n>2,$ the generic Lie symmetry vector is%
\begin{equation}
X=\xi^{i}\left(  x^{k}\right)  \partial_{i}+\left(  \frac{2-n}{2}\psi\left(
x^{k}\right)  u+a_{0}u+b\left(  x^{k}\right)  \right)  \partial_{u}%
\end{equation}
where $\xi^{i}$ is a CKV with conformal factor $\psi\left(  x^{k}\right)
$,$~b\left(  x^{k}\right)  $ is a solution of (\ref{lsk.02})~and the following
condition is satisfied%
\begin{equation}
\xi^{k}V_{,k}+2\psi V-\frac{2-n}{2}\Delta\psi=0.\label{KG.Eq22}%
\end{equation}
b) for $n=2,$ the generic Lie symmetry vector is
\begin{equation}
X=\xi^{i}\left(  x^{k}\right)  \partial_{i}+\left(  a_{0}u+b\left(
x^{k}\right)  \right)  \partial_{u}%
\end{equation}
where $\xi^{i}$ is a CKV with conformal factor $\psi\left(  x^{k}\right)
$,$~b\left(  x^{k}\right)  $ is a solution of (\ref{lsk.02}) and the following
condition is satisfied%
\begin{equation}
\xi^{k}V_{,k}+2\psi V=0.
\end{equation}

\end{proposition}

Let the potential $V\left(  x^{k}\right)  $ be
\begin{equation}
V\left(  x^{k}\right)  =-\frac{n-2}{4\left(  n-1\right)  }R\left(
x^{k}\right)  +\bar{V}\left(  x^{k}\right) \label{Ct.00.1}%
\end{equation}
where $R\left(  x^{k}\right)  $ is the Ricci scalar of the metric $g_{ij}$. In
that case, equation (\ref{lsk.02}) is called the conformal Klein Gordon
equation (or Yamabe equation). The conformal Klein Gordon equation~plays a
central role in the study of a conformal class of metrics by means of the
Yamabe invariant (see e.g. \cite{LieParker}).

Replacing $V\left(  x^{k}\right)  \ $from (\ref{Ct.00.1}) in the Lie symmetry
condition (\ref{KG.Eq22}), we have
\begin{equation}
\xi^{k}\bar{V}_{,k}+2\psi\bar{V}-\frac{n-2}{4\left(  n-1\right)  }\left(
\xi^{k}R_{,k}+2R\psi\right)  -\frac{2-n}{2}\Delta\psi=0.
\end{equation}
Because $\xi^{i}$ is a CKV with conformal factor $\psi$ it follows that
\cite{HallSteele}%
\begin{equation}
\xi^{k}R_{,k}=-2\psi R-2\left(  n-1\right)  \Delta\psi\label{Ct.02}%
\end{equation}
and this implies
\[
\left(  \frac{2-n}{2}\Delta\psi+\frac{n-2}{4\left(  n-1\right)  }\xi^{k}%
R_{,k}+\frac{n-2}{2\left(  n-1\right)  }\psi R\right)  =0.
\]

Therefore for the conformal Klein Gordon equation the symmetry condition
(\ref{KG.Eq22}) of theorem \ref{KG1} takes the final form
\begin{equation}
\xi^{k}\bar{V}_{,k}+2\psi\bar{V}=0.\label{Ct.03}%
\end{equation}

In the case where $V\left(  x^{k}\right)  =0$, the Klein Gordon equation
becomes the Laplace equation $~\Delta u=0$ and if $V\left(  x^{k}\right)
=-\frac{n-2}{4\left(  n-1\right)  }R\left(  x^{k}\right)  $ we have the
conformal Laplace equation
\begin{equation}
\Delta u+\frac{n-2}{4\left(  n-1\right)  }R\left(  x^{k}\right)
u=0.\label{CKG.01A}%
\end{equation}
A direct result, which arises from Proposition \ref{KG1} and the symmetry
condition (\ref{Ct.03}) for the Yamabe equation, is that, if $V^{n}$ is an $n
$ ($n>2$) dimensional Riemannian space, then, if the Laplace equation in
$V^{n}$ is invariant under a Lie group $G_{L},$~then, $\ G_{L}$ is a subgroup
\ of $\bar{G}_{L_{C}}$, i.e.~$G_{L}\subseteq\bar{G}_{LC}$ where $\bar{G}_{LC}$
is a Lie group which leaves invariant the conformal Laplace equation. The Lie
algebras $G_{L},\bar{G}_{LC}$ are identical if $V^{n}$ does not admit proper
CKVs or if all the conformal factors of the CKVs of $V^{n}~$are solutions of
the Laplace equation. Moreover, if $V^{n}$ is a conformally flat spacetime
then, the conformal Laplace equation (\ref{CKG.01A}) admits a Lie algebra of
$\frac{1}{2}\left(  n+1\right)  \left(  n+2\right)  +2$ \ dimension. For
instance, the Laplace equation in the three dimensional sphere\footnote{The
three dimensional sphere is a conformally flat space and it is maximally
symmetric.} $S^{3}$ admits eight Lie point symmetries \cite{Freire2010} (the
six KVs of $S^{3}$ plus the trivial symmetries $u\partial_{u},~b\left(
x^{k}\right)  \partial_{u}$) while, on the contrary, the conformal Laplace
equation admits twelve Lie symmetries, i.e. the conformal algebra of $S^{3}%
~$plus the trivial symmetries.

The Lie point symmetries of Schr\"{o}dinger equation follow from the elements
of the homothetic algebra of the metric~$g_{ij}$, whereas the Lie point
symmetries of the Klein Gordon equation are generated by the CKVs of the
metric which defines the Laplace operator. Consequently a relation between the
Noether point symmetries of the classical Lagrangian (\ref{CLN.05}) and the
Lie point symmetries of equations (\ref{lsk.01}) and (\ref{lsk.02}) is expected.

From the form of the symmetry vectors and the symmetry conditions for the
Schr\"{o}dinger equation (for details see \cite{2DPot}) we have the following result.

\begin{proposition}
\label{SymKGHJ11}If a KV/HV of the metric $g_{ij}$ produces a Lie point
symmetry of the Schr\"{o}dinger equation (\ref{lsk.01}), then generates a
Noether point symmetry for the Lagrangian (\ref{CLN.05}) in the space with
metric $g_{ij}$ and potential $V\left(  x^{k}\right)  .~$The reverse is also true.
\end{proposition}

Furthermore for the Klein Gordon equation we have the following proposition
which relates the Lie point symmetries of (\ref{lsk.02}) with the Noether
point symmetries of the conformal Lagrangian.

\begin{proposition}
\label{SymKGHJ}The Lie point symmetries of the Klein Gordon equation
(\ref{lsk.02}) for the metric $g_{ij}$ which defines the Laplace operator are
related to the Noether point symmetries of the classical Lagrangian
(\ref{CLN.05}) for the same metric and the same potential as follows

a) If a KV or a HV of the metric $g_{ij}$ generates a Lie point symmetry of
the Klein Gordon equation (\ref{lsk.02}), then it also produces a Noether
point symmetry for the classical Lagrangian with gauge function a constant.

b)~If a special CKV or a proper CKV of the metric $g_{ij}$ generates a Lie
point symmetry of the Klein Gordon equation (\ref{lsk.02}), then it also
generates a Noether point symmetry with gauge function a constant for the
conformally related\ Lagrangian provided there exists a conformal factor
$N\left(  x^{k}\right)  $ such that the CKV becomes a KV or a HV of the
conformal related metric $\bar{g}_{ij}$.
\end{proposition}

\section{Symmetries of the Lagrangian with non constant gauge function and the
Klein Gordon equation}

\label{Symmetries of the Lagrangian with non constant gauge function}

In the previous considerations we have shown that the Lie point symmetries of
the Klein Gordon equation induce Noether point symmetries for the classical
Lagrangian if the gauge function is a constant. In this section we investigate
the case when the induced Noether symmetry has a gauge function which is not a
constant. As we shall show, in this case the induced Noether symmetry comes
from a generalized Lie symmetry of the Klein Gordon equation. In particular we
study the case where the gradient KV and the gradient HV of the kinetic metric
generates Noether point symmetries for the classical Lagrangian.

\subsection{The oscillator potential}

Consider the Lagrangian of a particle moving in a decomposable Riemannian
space (i.e. the space admits a gradient KV \cite{TNA})
\begin{equation}
L=\frac{1}{2}\left(  \dot{x}^{2}+h_{AB}\left(  y^{C}\right)  \dot{y}^{A}%
\dot{y}^{B}\right)  +\mu^{2}x+F\left(  y^{C}\right) \label{CR.00}%
\end{equation}

For general functions $h_{AB}\left(  y^{C}\right)  ,~F\left(  y^{C}\right)  $,
Lagrangian (\ref{CR.00}) admits the autonomous Noether symmetry $\partial_{t}$
\ and two Noether symmetries $e^{\pm\mu t}\partial_{x}~$\cite{Ermakov} which
are due to the gradient KV $\partial_{x}.$. The corresponding Noether
integrals are%
\begin{equation}
I_{\pm}=e^{\pm\mu t}\dot{x}\mp\mu e^{\pm\mu t}x.
\end{equation}

It is straightforward to show that the combined integral $I_{0}=I_{+}I_{-}$ is
time independent and equals%
\begin{equation}
I_{0}=\dot{x}^{2}-\mu^{2}x^{2}.
\end{equation}

The Laplace Klein Gordon equation defined by the same "kinetic" metric and
potential is%
\begin{equation}
u_{xx}+h^{AB}u_{A}u_{B}-\Gamma^{A}u_{A}-\mu^{2}x^{2}u-F\left(  y^{C}\right)
u=0.\label{CR.05}%
\end{equation}
This equation does not admit a Lie point symmetry for general $h_{AB}\left(
y^{C}\right)  ,$ $F\left(  y^{C}\right)  .$ However it is separable with
respect to $x$ in the sense that the solution can be written in the form
$u\left(  x,y^{A}\right)  =w\left(  x\right)  S\left(  y^{A}\right)  .$ This
implies that the operator $\hat{I}=D_{x}D_{x}-\mu^{2}x^{2}-I_{0}$ satisfies
$\hat{I}u=0$ which means that the Klein Gordon equation (\ref{CR.05})
possesses a Lie B\"{a}cklund symmetry \cite{Miller,Kamr} with generating
vector $\bar{X}=\left(  u_{xx}-\mu^{2}x^{2}\right)  \partial_{u}$.$~$

Concerning the conformal Klein Gordon equation
\begin{equation}
u_{xx}+h^{AB}u_{A}u_{B}-\Gamma^{A}u_{A}+\frac{n-2}{4\left(  n-1\right)
}Ru-\mu^{2}x^{2}u-2\bar{F}\left(  y^{C}\right)  u=0\label{CR.06}%
\end{equation}
because for a KV $\xi^{a}$ we have $L_{\xi}R=0$ \cite{HallSteele} hence
$R=R\left(  y^{C}\right)  ,$ equation (\ref{CR.06}) is written in the form of
the Laplace Klein Gordon equation with $F\left(  y^{C}\right)  =2\bar
{F}\left(  y^{C}\right)  -\frac{n-2}{4\left(  n-1\right)  }R\left(
y^{C}\right)  $ and the previous results apply.

\subsection{The Ermakov potential}

We assume now that the classical Lagrangian (\ref{CLN.05}) admits as Noether
symmetries the $sl\left(  2,R\right)  $ Lie algebra $\left\{  \partial
_{t},\frac{1}{\mu}e^{\pm2\mu t}\partial_{t}\pm e^{\pm2\mu t}r\partial
_{r}\right\}  $. The Lagrangian is
\begin{equation}
L=\frac{1}{2}\left(  \dot{r}^{2}+r^{2}h_{AB}\left(  y^{C}\right)  \dot{y}%
^{A}\dot{y}^{B}\right)  +\frac{1}{2}\mu^{2}r^{2}-\frac{F\left(  y^{C}\right)
}{r^{2}}.\label{CR.07}%
\end{equation}

The corresponding Noether integrals are the Hamiltonian $h$ and%
\begin{align}
I_{+}  & =\frac{h}{\mu}e^{2\mu t}-e^{2\mu s}r\dot{r}+\mu e^{2\mu t}%
r^{2}\label{GERSN.5}\\
I_{-}  & =\frac{h}{\mu}e^{-2\mu t}+e^{-2\mu t}r\dot{r}+\mu e^{-2\mu t}%
r^{2}.\label{GERSN.6}%
\end{align}

From the Noether integrals (\ref{GERSN.5}),(\ref{GERSN.6}) and the Hamiltonian
$h$ of (\ref{CR.07}) we construct the time independent first integral
$\Phi_{0}=h^{2}-I_{+}I_{-}$ which is%
\begin{equation}
\Phi_{0}=r^{4}h_{DB}\dot{y}^{A}\dot{y}^{B}+2F\left(  y^{C}\right)
.\label{Er22}%
\end{equation}
This is the well known Ermakov invariant \cite{Lewis,LMO}. An alternative way
to construct the Ermakov invariant (\ref{Er22}) is to use dynamical Noether
symmetries \cite{Kalotas}. Indeed, one can show that the Lagrangian
(\ref{CR.07}) admits the dynamical Noether symmetry $X_{D}=K_{j}^{i}\dot
{x}^{j}\partial_{i}$ where $K^{ij}$ is a Killing tensor of the second rank
given by $K^{ij}=h^{AB}$.

The Laplace Klein Gordon equation defined by the Lagrangian (\ref{CR.07}) is%
\begin{equation}
u_{rr}+\frac{1}{r^{2}}h^{AB}u_{AB}+\frac{n-1}{r}u_{r}-\frac{1}{r^{2}}%
\Gamma^{A}u_{A}+\mu^{2}r^{2}+\frac{2}{r^{2}}F\left(  y^{C}\right)
=0.\label{CR.11}%
\end{equation}
This equation does not admit a Lie point symmetry. However it is separable in
the sense that $u\left(  r,y^{C}\right)  =w\left(  r\right)  S\left(
y^{C}\right)  .$ Then the operator%
\[
\hat{\Phi}=h^{AB}D_{A}D_{B}-\Gamma^{A}D_{A}+2F\left(  y^{C}\right)  -\Phi_{0}%
\]
satisfies the equation $\hat{\Phi}u=0$ which means that (\ref{CR.11}) admits
the Lie B\"{a}cklund\ symmetry with generator $\bar{X}=\left(  \hat{\Phi
}u\right)  \partial_{u}~$\cite{Miller,Kamr}.

Concerning the conformal Klein Gordon equation, the Ricci scalar of the metric
(\ref{CR.07}) and the HV satisfy the condition~ $L_{H}R+2R=0$
\cite{HallSteele}~that is $R=\frac{1}{r^{2}}\bar{R}\left(  y^{C}\right)  .$
Then, as in the case of the gradient KV, we absorb the term $\bar{R}\left(
y^{C}\right)  $ into the potential and we obtain the same results with the
Laplace Klein Gordon equation.

We conclude that although in the two cases considered above the Lie point
symmetries do not transfer from the classical to the "quantum" level the
generalized symmetries do transfer.

\section{Lie symmetry classification of Schr\"{o}dinger and Klein Gordon
equations in Euclidian space}

\label{Classification}

In \cite{2DPot,3Dpot} all two and three dimensional potentials have been
determined for which the corresponding Newtonian dynamical systems admit Lie
and Noether point symmetries. Using these results and propositions
\ref{SymKGHJ11},\ref{SymKGHJ} above we determine in this section all
potentials for which the Schr\"{o}dinger and Klein Gordon equation in
Euclidian 2d and 3d space admit Lie point symmetries.

The Schr\"{o}dinger equation in Euclidian space is%
\begin{equation}
\delta^{ij}u_{ij}+V\left(  x^{k}\right)  u=u_{t}.\label{SC1.01}%
\end{equation}
From proposition \ref{SymKGHJ11}, we have that the potentials for which the
Schr\"{o}dinger equation (\ref{SC1.01}) admits Lie point symmetries are the
same with the ones admitted by the classical Lagrangian. Therefore, the
results of \cite{2DPot,3Dpot} apply directly and give all potentials for which
the Schr\"{o}dinger equation (\ref{SC1.01}) admits at least one Lie point symmetry.

Concerning the Klein Gordon equation in flat space, that is the equation%
\begin{equation}
\delta^{ij}u_{ij}+V\left(  x^{k}\right)  u=0\label{LC.01}%
\end{equation}
from proposition \ref{SymKGHJ} we infer that equation (\ref{LC.01}) admits a
Lie point symmetry due to a HV/KV for the following potentials taken from the
corresponding Tables of \cite{2DPot,3Dpot}.\newline(a) Two dimensional: Table
13 (with $c=0$) and Table 14 (with $c=0)$ of \cite{2DPot}.\newline(b) Three
dimensional: Table 5, Table A1 (with $p=0)$ and Table A2 (with $p=0)$ of
\cite{3Dpot}. In Table \ref{2dkg} and Table \ref{3dkg} we give the potentials for the 2d and
the 3d case for which the Klein Gordon equation (\ref{LC.01})\ admits Lie
point symmetries.%

\begin{table}[ht]
\caption{The 2d (flat) Klein Gordon admits Lie symmetries from the
homothetic group}%
\begin{tabular}{@{}cccc@{}} \toprule
\textbf{Lie Symmetry} & $\mathbf{V}\left(  x,y\right)  $ & \textbf{Lie
Symmetry} & $\mathbf{V}\left(  x,y\right)  $\\ \colrule
$\partial_{x}$ & $f\left(  y\right)  $ & $\partial_{x}+b\partial_{y}$ &
$f\left(  y-bx\right)  $\\
$\partial_{y}$ & $f\left(  x\right)  $ & $\left(  a+y\right)  \partial
_{x}+\left(  b-x\right)  \partial_{y}$ & $f\left(  \frac{1}{2}\left(
x^{2}+y^{2}\right)  +ay-bx\right)  $\\
$y\partial_{x}-x\partial_{y}$ & $f\left(  r\right)  $ & $\left(  x+ay\right)
\partial_{x}+\left(  y-ax\right)  \partial_{y}$ & $r^{-2}~f\left(  \theta-a\ln
r\right)  $\\
$x\partial_{x}+y\partial_{y}$ & $x^{-2}f\left(  \frac{y}{x}\right)  $ &
$\left(  a+x\right)  \partial_{x}+\left(  b+y\right)  \partial_{y}$ &
$f\left(  \frac{b+x}{a+x}\right)  \left(  a+x\right)  ^{-2}$\\ \botrule
\end{tabular}
\label{2dkg}%
\end{table}%

As we have seen in section \ref{LiePointSchKG} the Lie point symmetries of the
Klein Gordon equation are generated from the conformal group of the space,
therefore in addition to the above we have to consider the admitted CKVs for
each case.%

\begin{table}[ht]
\caption{The 3d (flat) Klein Gordon admits Lie symmetries from the
homothetic group}
\begin{tabular}{@{}cccc@{}} \toprule
\textbf{Lie Symmetry} & $\mathbf{V(x,y,z)}$\\ \colrule
$a\partial_{\mu}+b\partial_{\nu}+c\partial_{\sigma}$ & $f\left(  x^{\nu}%
-\frac{b}{a}x^{\mu},x^{\sigma}-\frac{c}{a}x^{\mu}\right)  $\\
$a\partial_{\mu}+b\partial_{\nu}+c\left(  x_{\nu}\partial_{\mu}-x_{\mu
}\partial_{\nu}\right)  $ & ~$+f\left(  \frac{c}{2}r_{\left(  \mu\nu\right)
}-bx_{\mu}+ax_{\nu},x_{\sigma}\right)  $\\
$a\partial_{\mu}+b\partial_{\nu}+c\left(  x_{\sigma}\partial_{\mu}-x_{\mu
}\partial_{\sigma}\right)  $ & $+f\left(  x_{\nu}-\frac{1}{\left\vert
c\right\vert }\arctan\left(  \frac{\left\vert c\right\vert x_{\mu}}{\left\vert
a+cx_{\sigma}\right\vert }\right)  ,\frac{1}{2}r_{\left(  \mu\sigma\right)
}-\frac{a}{c}x_{\sigma}\right)  $\\
$a\partial_{\mu}+b\left(  x_{\nu}\partial_{\mu}-x_{\mu}\partial_{\nu}\right)
+$ & \\
$~~+c\left(  x_{\sigma}\partial_{\mu}-x_{\mu}\partial_{\sigma}\right)  $ &
$f\left(  x_{\mu}^{2}+x_{\nu}^{2}\left(  1-\frac{c^{2}}{b^{2}}\right)
+\left(  \frac{2a}{b}+\frac{2c}{b}x_{\sigma}\right)  x_{\nu},x_{\sigma}%
-\frac{c}{b}x_{\nu}\right)  $\\
$so\left(  3\right)  $ linear combination & $~F\left(  R,b\tan\theta\sin
\phi+c\cos\phi-aM_{1}\right)  $\\
$a\partial_{\mu}+b\theta_{\left(  \nu\sigma\right)  }\partial_{\theta_{\left(
\nu\sigma\right)  }}+cR\partial_{R}$ & $\frac{1}{r_{\left(  \nu\sigma\right)
}^{2}}f\left(  \theta_{\left(  \nu\sigma\right)  }-\frac{b}{c}\ln r_{\left(
\nu\sigma\right)  },\frac{a+cx_{\mu}}{cr_{\left(  \nu\sigma\right)  }}\right)
$\\
$\left(  a\partial_{\mu}+b\partial_{\nu}+c\partial_{\sigma}+lR\partial
_{R}\right)  $ & $\frac{1}{\left(  a+lx_{\mu}\right)  ^{2}}f\left(
\frac{b+lx_{\nu}}{l\left(  a+lx_{\mu}\right)  },\frac{c+lx_{\sigma}}{l\left(
a+lx_{\mu}\right)  }\right)  $\\\botrule
\end{tabular}
\label{3dkg}
\end{table}%

\subsection{The two dimensional case}

We recall that the conformal algebra of a two dimensional space is infinite
dimensional \cite{Bela} and in coordinates with line element $ds^{2}=2dwdz$
are given by the vectors $X=F\left(  w\right)  \partial_{w}+G\left(  w\right)
\partial_{w}$ with conformal factor $\psi=\frac{1}{2}\left(  F_{,w}%
+G_{,z}\right)  $. In the coordinates $\left(  w,z\right)  $ the 2d Klein
Gordon equation (\ref{LC.01}) is%
\[
u_{wz}+V\left(  w,z\right)  u=0.
\]
The Lie symmetry condition (\ref{KG.Eq22}) becomes%
\begin{equation}
\left(  FV\right)  _{,w}+\left(  GV\right)  _{,z}=0
\end{equation}
from which follows that there are infinite many potentials for which the 2d
Klein Gordon equation admits Lie point symmetries.

\subsection{The three dimensional case}

The 3d Euclidian space admits the three sp.CKVs%
\[
K_{C}^{\mu}=\frac{1}{2}\left(  \left(  x^{\mu}\right)  ^{2}-\left(  \left(
x^{\sigma}\right)  ^{2}+\left(  x^{\nu}\right)  ^{2}\right)  \right)
\partial_{i}+x^{\mu}x^{\nu}\partial_{x}+x^{\mu}x^{\nu}\partial_{z}%
~,~\mu,\sigma,\nu=1,2,3
\]
with corresponding conformal factor $\psi_{C}=x^{\mu}$.

From the symmetry condition (\ref{KG.Eq22}) of Proposition \ref{KG1} follows
that a sp.CKV generates the Lie point symmetry $X=K_{C}^{\mu}-\frac{1}%
{2}x^{\mu}u\partial_{u}$ for the 3d Klein Gordon (\ref{LC.01}) only for the
potential
\[
V\left(  x,y,z\right)  =\frac{1}{\left(  x^{\sigma}\right)  ^{2}}F\left(
\frac{x^{\nu}}{x^{\sigma}},\frac{\delta_{\kappa\lambda}x^{\kappa}x^{\lambda}%
}{x^{\sigma}}\right)  .
\]

One is possible to continue with the linear combinations of these symmetry
vectors and deduce all cases that the 3d Klein Gordon equation admits a Lie
point symmetry (see \cite{2DPot,3Dpot}).

These results hold for both the Klein Gordon and the conformal Klein Gordon
equation. In particular it can be shown that the results remain valid for the
conformal Klein Gordon equation provided the metric defining the conformal
Laplacian is conformally flat.

\section{Spherically symmetric space-time}

\label{StaticSS}

In this section we consider the Lie point symmetries of the Klein Gordon
equation in a non-flat space and in particular in the static spherically
symmetric empty space-time, that is the exterior Schwarzschild solution, given
by the metric ($\tau$ is the radial coordinate)
\begin{equation}
ds^{2}=-a^{2}\left(  \tau\right)  dt^{2}+d\tau^{2}+b^{2}\left(  \tau\right)
\left(  d\theta^{2}+\sin^{2}\theta d\phi^{2}\right)  .\label{SC}%
\end{equation}
The Lagrangian of Einstein field equations for this space-time is
\cite{VakB,Christ}%
\begin{equation}
L=2ab^{\prime2}+4ba^{\prime}b^{\prime}+2a\label{SC.01}%
\end{equation}
where $"^{\prime}"~$means derivative with respect to the radius $\tau$. If we
see the Lagrangian (\ref{SC.01}) as a dynamical system in the space of
variables $\left\{  a,b\right\}  $, then this system is "autonomous" hence
admits the Noether point symmetry $\partial_{\tau}$ with corresponding Noether
integral\footnote{$h$ \emph{is not} the "energy" because the coordinate is the
radial distance not the time} the "Hamiltonian" $h=$constant. We compute
\[
h=2ab^{\prime2}+4ba^{\prime}b^{\prime}-2a.
\]

It is straightforward to show that $h=\frac{2}{a}G_{1}^{1}$ where $G_{1}^{1}$
is the Einstein tensor. Because the space is empty from Einstein's equations
follows that $h=0.$ Euler-Lagrange equations are%
\[
a^{\prime\prime}-\frac{a}{2b^{2}}b^{\prime2}+\frac{1}{b}a^{\prime}b^{\prime
}+\frac{1}{2}\frac{a}{b^{2}}=0
\]%
\[
b^{\prime\prime}+\frac{1}{2b}b^{\prime2}-\frac{1}{2b}=0.
\]

We end up with a system of three equations whose solution will give the
functions $a(\tau),b(\tau).$ However it is found that the solution of the
system does not give these functions in the well known closed form. This is
due to the Lagrangian we have considered. Indeed as we shall show below it is
possible to find the solution in closed form by considering a Lagrangian
conformally related to the Lagrangian (\ref{SC.01}). It is straightforward to
show that the Lagrangian (\ref{SC.01}) admits the Noether symmetry
\[
X_{1}=2\tau\partial_{\tau}+H
\]
where $H=-2a\partial_{a}+2b\partial_{b}$ is a non-gradient homothetic vector
of the two dimensional kinetic metric
\begin{equation}
d\bar{s}^{2}=2adb^{2}+4bdadb\label{SC.01a}%
\end{equation}
defined from the Lagrangian (\ref{SC.01}). Consider the Klein Gordon equation
defined by the metric (\ref{SC.01a}) with potential $V\left(  a,b\right)
=2a$, that is,%
\begin{equation}
-\frac{a}{4b^{2}}u_{aa}+\frac{1}{2b}u_{ab}-\frac{1}{4b^{2}}u_{a}%
-2au=0.\label{SC.02}%
\end{equation}
Equation (\ref{SC.02}) admits as Lie point symmetries the vectors
\cite{Christ}%
\[
u\partial_{u}~~,~b\left(  a,b\right)  \partial_{u}%
\]%
\[
H=-2a\partial_{a}+2b\partial_{b}~~,~~X_{2}=\frac{1}{ab}\partial_{a}%
~~,~X_{3}=\frac{a}{2b}\partial_{a}-\partial_{b}%
\]
where the vectors $X_{2},X_{3}$ are proper CKVs of the two dimensional metric
(\ref{SC.01a}).

By applying the Lie invariants it is possible to find solutions of the Klein
Gordon equation (\ref{SC.02}) which are invariant with respect to one of the
admitted Lie point symmetries.

For example let us consider the Lie point symmetry $H_{u}=H-2cu\partial_{u}. $
The zero order invariants $H_{u}$ are ~$w=ab~,~u=a^{c}\Phi\left(  w\right)  .$
Replacing in (\ref{SC.02}) we find the solution%

\[
u\left(  a,b\right)  =a^{c}\left[  c_{1}I_{c}^{B}\left(  2\sqrt{2}ab\right)
+c_{2}K_{c}^{B}\left(  2\sqrt{2}ab\right)  \right]
\]
where $I^{B},K^{B}$ are the Bessel modified functions \cite{Christ}. Working
similarly for the Lie point symmetry $H+eX_{2}-cu\partial_{u}$ we find the
solution%
\[
u\left(  a,b\right)  =\left(  a^{2}-eb^{-1}\right)  ^{\frac{c}{4}}\left[
c_{1}I_{-\frac{c}{2}}\left(  2\sqrt{2b\left(  a^{2}b-e\right)  }\right)
+c_{2}K_{\frac{c}{2}}^{B}\left(  2\sqrt{2b\left(  a^{2}b-e\right)  }\right)
\right]  .
\]

One can find more solutions using linear combinations of these Lie symmetries.

Following proposition \ref{SymKGHJ} we look for a conformal metric for which
one of the CKVs $X_{2},X_{3}$ becomes a KV\ and write the corresponding
conformally related Lagrangian which admits this CKV\ as a Noether symmetry.
We consider first the vector $X_{2}$ and the conformally related metric
\begin{equation}
ds^{2}=N^{2}\left(  4adb^{2}+8bda~db\right) \label{M1}%
\end{equation}
where $N\left(  a,b\right)  =g\left(  b\right)  \sqrt{a}$ and $g\left(
b\right)  $ is an arbitrary function of its argument$.$ It is easy to show
that the vector $X_{2}$ is a KV of this metric hence a Noether symmetry for
the family of conformally related Lagrangians
\begin{equation}
\bar{L}=N^{2}\left[  2a\left(  \frac{db}{dr}\right)  ^{2}+4b\left(  \frac
{da}{dr}\right)  \left(  \frac{db}{dr}\right)  \right]  +\frac{2a}{N^{2}%
}\label{M2}%
\end{equation}
where we have considered the coordinate transformation $d\tau=\frac{dr}%
{N^{2}\left(  a,b\right)  }$. The Noether function for this Noether symmetry
is the "Hamiltonian" of the Lagrangian (\ref{M2}). This constant and the two
Lagrange equations for the "generalized" coordinates $a,b$ provide a system of
differential equations which will give the functions $a(\tau),b(\tau).$

We consider $g\left(  b\right)  =g_{0}=$constant, that is from the family of
Lagrangians (\ref{M2}) we take the Lagrangian%
\begin{equation}
L=g_{0}^{2}\left[  2a^{2}\left(  \frac{db}{dr}\right)  ^{2}+4ab\left(
\frac{da}{dr}\right)  \left(  \frac{db}{dr}\right)  \right]  +\frac{2}%
{g_{0}^{2}}.\label{M2A}%
\end{equation}
For this Lagrangian the field equations are the "Hamiltonian" of the
Lagrangian (\ref{M2A})
\begin{equation}
a^{2}\left(  \frac{db}{dr}\right)  ^{2}+2b^{2}\left(  \frac{da}{dr}\right)
\left(  \frac{db}{dr}\right)  -V_{0}=0\label{M3}%
\end{equation}
and the Euler Lagrange equations of (\ref{M2A}) with respect to the variables
$a,b~$ \
\begin{equation}
\frac{d^{2}a}{dr^{2}}+\frac{1}{a^{2}}\left(  \frac{da}{dr}\right)  ^{2}%
+\frac{2}{b}\left(  \frac{da}{dr}\right)  \left(  \frac{db}{dr}\right)
=0\label{M4}%
\end{equation}

\begin{equation}
\frac{d^{2}b}{dr^{2}}=0\label{M5}%
\end{equation}
where we have set $V_{0}=g_{0}^{-4}$.

The solution of the system of equations (\ref{M3})-(\ref{M5}) is
\[
b\left(  r\right)  =b_{1}r+b_{2}~~,~~~a^{2}\left(  r\right)  =\frac
{V_{0}r+2a_{1}b_{1}}{2b_{1}\left(  b_{1}r+b_{2}\right)  }.
\]
Under the linear transformation $b_{1}r=b_{1}R-\frac{b_{2}}{b_{1}}$ and if we
set $V_{0}=2\left(  b_{1}\right)  ^{2}~,~a_{1}=-2m+b_{2}$, we obtain the
exterior Schwarzschild solution in the standard coordinates%
\begin{equation}
ds^{2}=-\left(  1-\frac{2m}{R}\right)  dt^{2}+\left(  1-\frac{2m}{R}\right)
^{-1}dR^{2}+R^{2}\left(  d\theta^{2}+\sin^{2}\theta d\phi^{2}\right)
\label{M6}%
\end{equation}

The choice of the function $g\left(  b\right)  $ is essentially a choice of
the coordinate system. Obviously the final solution must always be the
exterior Schwarzschild solution. Working similarly we find that $X_{3}$
becomes a KV for the conformal metric (\ref{M1}) if $N_{3}\left(  a,b\right)
=f\left(  a^{2}b\right)  \sqrt{a}$, and generates a Noether point symmetry for
the conformal Lagrangian (\ref{M2}) (with $N_{3}$ in the place of $N)\ $\ and
from this point we continue as above.

\section{Conclusion}

\label{conclusion}

We have determined the Lie point symmetries of the Schr\"{o}dinger equation
and the Klein Gordon equation in a general Riemannian space. It has been shown
that these symmetries are related with the homothetic algebra and the
conformal algebra of the metric of the space. Furthermore these symmetries
have been related to the Noether point symmetries of the classical Lagrangian
for which the metric $g_{ij}$ is the kinematic metric. More precisely for the
Schr\"{o}dinger equation it has been shown that if a KV/HV of the metric
$g_{ij}$ produces a Lie point symmetry for the Schr\"{o}dinger equation then
produces a Noether point symmetry for the classical Lagrangian in the space
with metric $g_{ij}$ and potential $V(x^{k})$. For the Klein Gordon equation
the situation is different. That is the Lie point symmetries of the Klein
Gordon are generated from the conformal algebra of the metric $g_{ij}.$ The
KVs and the HV of this group produce a Noether symmetry of the classical
Lagrangian with a constant gauge function, however the proper CKVs produce a
Noether symmetry of the conformal Lagrangian if there exists a conformal
factor $N\left(  x^{k}\right)  $ such that the CKV becomes a KV/HV of $g_{ij}$.

{We have applied these general results to two cases of practical interest. The
classification of all potentials in Euclidian 2d and 3d space for which the
Schr\"{o}dinger equation and the Klein Gordon equation admit a Lie point
symmetry and finally we have considered the Lie point symmetries of the Klein
Gordon equation in the static, spherically symmetric empty spacetime. In the
last case we have demonstrated the role of the Lie symmetries and that of the
conformal Lagrangians in the determination of the closed form solution of
Einstein equations. }

The knowledge of Lie point symmetries of Schr\"{o}dinger equation and the
Klein Gordon equation in a general Riemannian space makes possible the
determination of solutions of these equations which are invariant under a
given Lie point symmetry. In addition the Lie point symmetries are used in
Quantum Cosmology \cite{CAP94,CKP,VakF,VF12,CapHam} in order to determine the
form of solutions of the Wheeler--DeWitt equation in a given Riemannian space.

\end{document}